# Wave-packet rectification in nonlinear electronic systems: A tunable Aharonov-Bohm diode

Yunyun Li[1], Jun Zhou[1*], Fabio Marchesoni[1,2] & Baowen Li[1,3]

[1]Center for Phononics and Thermal Energy Science, School of Physics Science and Engineering, Tongji University, Shanghai 200092, People's Republic of China

[2]Department of Physics, University of Camerino, I-62032 Camerino, Italy

[3]Department of Physics and Center for Computational Science and Engineering, National University of Singapore, Singapore 117456, Republic of Singapore

*Correspondence and requests for materials should be addressed to:
zhoujunzhou@tongji.edu.cn (J. Zhou)

**Abstract**

Rectification of electron wave-packets propagating along a quasi-one dimensional chain is commonly achieved via the simultaneous action of nonlinearity and longitudinal asymmetry, both confined to a limited portion of the chain termed wave diode. However, it is conceivable that, in the presence of an external magnetic field, spatial asymmetry perpendicular to the direction of propagation suffices to ensure rectification. This is the case of a nonlinear ring-shaped lattice with different upper and lower halves (diode), which is attached to two elastic chains (leads). The resulting device is mirror symmetric with respect to the ring vertical axis, but mirror asymmetric with respect to the chain direction. Wave propagation along the two diode paths can be modeled for simplicity by a discrete Schrödinger equation with cubic nonlinearities. Numerical simulations demonstrate that, thanks to the Aharonov-Bohm effect, such a diode can be operated by tuning the magnetic flux across the ring.

**Introduction**

The design of current rectifiers, also known as diodes, has spawned widespread interest in many branches of physics. The analogs of electronic diodes, such as thermal diodes[1-5], optical diode[6,7] and acoustic diodes[8,9], have been theoretically proposed to effectively manipulate energy transport and wave propagation. Thanks to recent advances in nanotechnology, prototypes of some of them have indeed been realized[10,11]. The operation of this class of quantum devices requires at least two ingredients, namely, nonlinearity and spatial asymmetry along the direction of propagation[12] (longitudinal asymmetry). A similar conclusion was reached in the classical theory of Feynman's ratchets and Brownian motors, where autonomous transport of neutral particles and other small objects in nonlinear asymmetric devices is sustained by ambient fluctuations[13].

In electronic systems the interaction between electrons and lattice vibrations, or phonons, is responsible for appreciable nonlinear transport effects[14-16]. This scenario applies, for instance, to conductive polymers, where solitons play a fundamental role in the charge-transfer doping mechanism[17]; to macromolecules such as DNA, where the nonlinear electron-phonon coupling is likely to play an important role[18-21]. In this context the dynamics of an electron wave-packet subject to nonlinear electron-phonon interactions was conveniently modeled in terms of discrete nonlinear Schrödinger equations (DNLSE)[22]. The DNLSE's have also been employed to describe the propagation of longitudinal Block waves in layered or phononic crystals[23], light through networks of coupled optical waveguides in Kerr media[24] and interacting cold atoms on an optical lattice[25].

A composite quasi-one dimensional nonlinear device inserted along a transmission line operates as wave diode when its transmission coefficients depend on the direction of the incoming wave. The required mechanism of longitudinal symmetry breaking is typically established by grading the composition and geometry of the diode components[26,27]. Moreover, for practical purposes any such device must be tunable, which raises the question of how to control the diode output by means of an external signal.

In this paper we tackle both issues of asymmetry and control, by considering a novel diode topology. A sketch of the device we propose is shown in Figure 1, where a ring (diode) is connected to two chains (leads) parallel to one of its diameters. Both the chains and the ring are modeled respectively as elastic and nonlinear discrete lattices. The upper and lower halves of the ring have different propagation properties. With this design the left-right mirror (or longitudinal) symmetry of the device is preserved, whereas its upside-down (or transverse) symmetry is broken. Under the operating conditions discussed in the earlier literature, such a device would be of no use as a wave rectifier[26]. However, this is no longer true after a magnetic field perpendicular to the ring plane is switched on. Owing to the nonlinear nature of the ring, the difference between the Aharonov-Bohm (AB) phases electrons acquire along the upper and lower paths depends on the side they enter the diode. We prove that, as a consequence, the left-right and right-left transmission coefficients across the AB ring can be different and the relevant rectification factor easily controlled by tuning the magnetic field.

**Results**

**Model**

The nonlinear propagation properties of the ring were modeled by the following DNSLE for the electron wave function $\psi_n$[28]

$$i\dot{\psi}_n = \varepsilon_n \psi_n + g_n |\psi_n|^2 \psi_n - \sum_m t_{n+m,n} \psi_{n+m}. \tag{1}$$

The sites of the left (right) chain are labeled from $-N_p$ to $-N_l-1$ (from $N_l+1$ to $N_p$); the sites of the upper (lower) ring half are labeled from $(-N_l+1)$ to $(N_l-1)$ [from $(-N_l+1)'$ to $(N_l-1)'$], and the ring-chain junction sites are $\pm N_l$. $\varepsilon_n$ represents the on-site energy and the constant $g_n$ is meant to reproduce the local nonlinear electron-phonon coupling. In particular we set: $\varepsilon_n = g_n = 0$ on the elastic chains; $\varepsilon_n = \varepsilon_\pm = \varepsilon_0(1\pm\delta_\varepsilon)$, $g_n = g_\pm = g_0(1\pm\delta_g)$ in the upper (lower) ring sides; and, finally, $\varepsilon_{\pm N_l} = \varepsilon_0$ and $g_{\pm N_l} = g_0$ at the junction sites. The parameters $\delta_\varepsilon$ and $\delta_g$ were used to control the asymmetry between the upper and lower halves of the ring, while $g_0$ was tuned to vary its overall nonlinearity.

In the last term of equation (1) the summation over m was restricted to hopping between nearest neighbors, for which $t_{n+m,n}=1$. All sites are coupled to two nearest neighbors except the junction sites with $n=\pm N_l$, which have three nearest neighbors with $n+m=\pm(N_l-1)$, $\pm(N_l-1)'$, and $\pm(N_l+1)$. The hopping rates between all other sites were set to zero, so that there is no interaction between the upper and the lower side of the ring.

The effect of the magnetic field on a charge q hopping along the AB ring was accounted for by means of the so-called "Peierls substitution"[29,30],

$$t_{n+m,n} \rightarrow t_{n+m,n} \exp\left(\pm i \frac{2\pi}{N_R} \frac{\Phi}{\Phi_0}\right), \qquad (2)$$

where $N_R=4N_I$ is the number of ring sites, $\Phi$ the magnetic flux connected to the ring, $\Phi_0=h/q$ the quantum of magnetic flux and $\pm$ stand for the positive or negative orientation of the n→n+m hop. The phase modulation in equation (2) is restricted to the hopping rates between ring sites.

We further notice that for our choice of $t_{n+m,n}$ equation (1) can be easily derived from the Hamiltonian $H[\psi] = \sum_n \left(V[\psi_n] - (\psi_{n+1}\psi_n^* + \psi_{n+1}^*\psi_n)\right)$, where the quartic on-site potential $V[\psi_n] = \varepsilon_n|\psi_n|^2 + g_n|\psi_n|^4/2$ is nonzero only inside the ring. Accordingly, both the energy and the norm of the propagating wave-packet are conserved, as expected for electrons propagating in a non-dissipative medium. In our simulations we focused on the case of incident Gaussian wave-packets,

$$\psi_n(0) = I_0 \exp\left(-\frac{(n-n_0)^2}{4\sigma_0^2} - ik_0 n\right), \qquad (3)$$

with relatively small wavenumbers, $\lambda_0=2\pi/k_0$, and large width $\sigma_0$, that is, $\lambda_0 \ll N_R$ and $\sigma_0 \gg N_R$. Along an elastic chain ($\varepsilon=g=0$) with $t_{n+m,n}=1$ for $m=\pm 1$ and $t_{n+m,n}=0$ otherwise, the group velocity of the wave-packet[27,28] is $v_p=2\sin k_0$.

**Nonlinearity**

The transmission coefficients $T_{LR}$ (left-right) and $T_{RL}$ (right-left) have been computed as the ratio of the norm of the wave-packet transmitted, respectively, to the right and left chain, to the (conserved) total initial norm (see Methods for computational details). In Figure 2a $T_{LR}$ and $T_{RL}$ have been plotted as functions of the magnetic flux, $\Phi$, for different values of the $V[\psi_n]$ parameters $\alpha=(\varepsilon,g)$ on the ring sites. Here, $\varepsilon$ is the same on both sides of the ring, $\delta_\varepsilon=0$, whereas g is not, $\delta_g=0.3$. The role of nonlinearity combined with asymmetry is apparent. Contrary to the symmetric and harmonic cases with $g_0=0$ but $\varepsilon_0\neq 0$ (dashed curves), the transmission coefficients across a nonlinear and upside-down asymmetric ring, $T_{LR}(\Phi)$ (solid, blue) and $T_{RL}(\Phi)$ (solid, red), do depend on the direction of propagation.

Qualitatively, this can be explained as an interference effect. Entering the ring from the left (right) at the junction site $-N_l$ ($+N_l$), the wave-packet splits into two wave-packets (not necessarily with the same norm), which then propagate respectively along the upper and lower path. The magnetic flux changes their phase with opposite sign. As $\sigma_0>>N_R$, when the two wave-packets recombine at the exit junction sites $N_l$ ($-N_l$), they interfere. For a given $\Phi$ their phase difference at recombination depends on the parameters of the upper and lower ring sides. As long as the on-site potential is harmonic, $g_0=0$, changing the direction of propagation only reverses the sign of the phase difference, so that $T_{LR}(\Phi)=T_{RL}(\Phi)$. For $g_0\neq 0$, however, site resonances are detuned differently for $k_0>0$ and $k_0<0$, which leads to different transmission coefficients[26].

The $\Phi$ dependence of the transmission coefficients in the harmonic case, $g_0=0$, is also remarkable. For $\varepsilon_0<2$ the curves $T_{LR,RL}(\Phi)$ are bimodal with symmetric maxima (close to 1) for $\varepsilon_0$-dependent values of the flux. On increasing $\varepsilon_0$ the two transmission peaks tend to merge into a lower single peak centered at $\Phi=0$. Note that in any case the transmission coefficients are appreciably larger than 0.5 (incoherent propagation regime) only in the central flux range, $-0.25<\Phi/\Phi_0<0.25$.

To further characterize the asymmetric transmission properties of our wave diode, in Figure 2b we also plotted the rectification factor $R_f$,

$$R_f = \frac{T_{LR}(k_0)-T_{RL}(-k_0)}{T_{LR}(k_0)+T_{RL}(-k_0)} \quad , \tag{4}$$

as a function of Φ. Its overall magnitude strongly depends on the nonlinearity constant $g_0$. On keeping the asymmetry parameters fixed, $δ_ε=0$ and $δ_g=0.3$, the amplitude of $R_f$ goes through a maximum for an optimal value of $g_0$, while $T_{LR}$ and $T_{RL}$ are drastically lowered at larger $g_0$ (not shown). Too strong a nonlinear coupling tends to suppress rectification.

Finally, we notice that inverting the sign of $g_0$ in Figure 2b does not appreciably affect the overall magnitude of the $R_f$ oscillations. However, the curve for $g_0=-1$ develops a strongly fluctuating Φ dependence, which hints at a chaotic behavior[26]. This feature has been detected any time $ε_0$ and $g_0$ had opposite sign, namely when the nonlinear on-site potential, $V[ψ_n]$, was bistable. Another example of rectification for negative nonlinearity constant is displayed in Figure 3.

**Flux as control parameter**

Figure 3 illustrates how the rectification power of the diode can be effectively controlled by tuning the magnetic field. Here we set Φ so as to maximize the difference between $T_{LR}$ and $T_{RL}$ (i.e. $R_f$) and visualized the time evolution of the incoming wave-packet. When the packet proceeds from the left, it passes through the ring undergoing minor distortion; its reflected image is barely detectable. Vice versa, when the wave-packed impacts the ring from the right, it gets almost entirely reflected, only a small fraction of it being transmitted. During the transmission process the norm of the traveling wave tends to accumulate on the ring sites to different degrees. In the lower panels of Figure 3 we plotted the probability spectral density (p.s.d., defined in Methods) on site n=0 for the two cases. The relevant p.s.d.'s peak respectively at zero frequency (wave not changing phase during the process), when the wave-packet gets transmitted, and at $2πv_p/N_R$ (wave shifting phase by π upon collision), when it gets reflected. A (nonlinear) analysis of the fine structure of the peak in Figure 3d rests outside the purpose of the present report.

Most remarkably, panels (a) and (b) of Figure 2 suggest that the symmetry relations $T_{LR}(\Phi)=T_{RL}(-\Phi)$ and, therefore, $R_f(-\Phi)=-R_f(\Phi)$ must hold for any choice of $\alpha$. Indeed, under the transformations $\alpha \to \varepsilon_\alpha \alpha$, $k_0 \to \varepsilon_k k_0$, $\Phi \to \varepsilon_\Phi \Phi$, with $\varepsilon_i=\pm 1$ and $i=\alpha, k, \Phi$, the dynamics of equation (1) with our choice of $t_{n+m,n}$ satisfies three distinct symmetry relations, which in compact form read (see Methods)

$$R_f(\varepsilon_\alpha \alpha, \varepsilon_k k_0, \varepsilon_\Phi \Phi) = \varepsilon_\alpha \varepsilon_k \varepsilon_\Phi \, R_f(\alpha, k_0, \Phi). \tag{5}$$

Here, $\varepsilon_\Phi=-1$ means that the sign of the AB phase (i.e., of the magnetic field) is reversed, $\varepsilon_k=-1$ indicates a change in the direction of the incident wave-packet (i.e., an exchange of the suffixes LR and RL) and, finally, $\varepsilon_\alpha=-1$ corresponds to turning the on-site potential, $V[\psi_n]$, upside-down. Our numerical data confirm all identities obtainable from equation (5).

**Asymmetry**

In panels (a) and (b) of Figure 4 we analyze the role of the second key ingredient of our rectifier, namely its upside-down mirror asymmetry. In panel (a) $\delta_g$ was increased at $\delta_\varepsilon=0$. The rectification factor in the domain $-0.25<\Phi/\Phi_0<0.25$ turned out to be rather sensitive to the asymmetry of the quartic term of $V[\psi_n]$. The slope of $R_f$ at $\Phi=0$ increases with $\delta_g$ up to a maximum. At even larger $\delta_g$ it changes sign and, as a consequence, a current inversion may occur as the curve $R_f(\Phi)$ oscillates around zero.

Asymmetry can be controlled also by varying $\delta_\varepsilon$ at $\delta_g=0$. As long as $g_0 \neq 0$, this suffices in principle to operate our wave diode, as shown in Figure 4b. However, contrary to the situation discussed above, here increasing $\delta_\varepsilon$ induces frequent sign changes in the slope of $R_f$ at $\Phi=0$, thus making an effective control of the diode operation more difficult.

Finally, we briefly discuss the coherence issue. For the two split wave-packets in the AB ring to interfere at the exit junction site, where they recombine, is necessary that the relative spatial delay accumulated by following respectively the upper and lower path, be much smaller than their width. In Figure 4c we display $T_{LR}$ as a function of the flux for different circumferences, $N_R$, of the AB ring. The $\Phi$ oscillations of the transmission coefficient get suppressed on increasing $N_R$ and, eventually, for $N_R \geq \sigma_0$ $T_{LR}$ fluctuates around its incoherent regime value, $T_{LR}=1/2$, which confirms the absence of interference between the two exiting wave packets.

**Discussion**

The wave diode we discussed in this report consists basically of an AB ring made of two halves with different nonlinear transport properties. Its operation principle should not be mistaken for the standard rectification mechanism demonstrated in the recent literature[13], because it has especially been designed to retain longitudinal mirror symmetry. Here, it is the combined action of *transverse* mirror asymmetry with respect to the direction of transmission and the AB phase acquired by the charges moving around ring, that causes a net *longitudinal* transmission current. The rectification power of the resulting wave diode is controlled by the magnetic flux threading through the ring. The special topology of our AB diode could be implemented ad hoc, for instance, by intertwining two different molecular wires or unzipping a DNA chain so as to create a one dimensional molecular filament with interspersed loops.

More interestingly, our wave-diode design suggests more direct applications to real electronic systems. Let us consider, for instance, a thin strip of a type II superconductor sandwiched between two strips of different conducting materials to form a three-layered long ribbon. On applying a constant magnetic field oriented parallel to the ribbon layers and perpendicular to the ribbon axis, one can generate one or more magnetic flux lines (fluxons), each of magnitude $\Phi_0$, which under appropriate fabrication conditions, can be trapped at special pinning sites in the central layer. The intensity of the applied magnetic field controls the fluxon density. Electrons moving in the outer ribbon layers are affected by the presence of such tightly sandwiched fluxons[32]. Indeed, on passing each flux line, they acquire an AB phase, which depends in sign and magnitude on which conducting strip they propagate along. On assuming that the material of at least one of the two conducting strips has nonlinear transport properties, the situation envisaged in our stylized tunable *Aharonov-Bohm diode* is thus actually realized.

Let us consider now an even simpler conducting ribbon by removing the inner superconducting strip. At the interface between the two remaining conducting strips, stresses induced, say, by mechanical or thermal deformation, can result in the nucleation of linear defects, dislocations, oriented perpendicular to the ribbon axis. Electrons moving past an edge dislocation accumulate an additional phase due to the distortion of the lattice structure that hosts the dislocation[33]. In this situation the role of the magnetic flux lines would be played by the dislocations – both fluxons and dislocations can be regarded as different species of linear defects. In this setup, however, tunability would more difficult as dislocation density and topology are hardly controllable factors. A similar geometry could also work as phonon rectifiers provided tht the edge dislocations are replaced by screw dislocations and the incident packet is made of acoustic waves[34].

**Methods**

**Simulations**. In our simulations the number of ring sites was $N_R=4N_l$, and the number chain sites on both sides of the ring $N_p = 500$. The wave-packets travelling to the left (right) were injected at time t=0 with width $\sigma_0=25$ and centered at $n_0=N_p/2$ (-$N_p/2$). At the endpoints of the right (left) chain, $n=\pm N_p$, we set $t_{m+n,n}=0$ for $m=\pm 1$ (reflecting boundaries). For the purposes of the present study these chains can be considered infinitely long. Due to their large width the distortion of the transmitted (reflected) wave-packets was negligible (see, e.g., Figure 3).

The explicit expressions employed to compute the transmission coefficients were

$$T_{LR} = \frac{\sum_{n=N_l}^{N_p}|\psi_n|^2}{\sum_{n=-N_p}^{-N_l}|\psi_n|^2}, \quad T_{RL} = \frac{\sum_{n=-N_p}^{-N_l}|\psi_n|^2}{\sum_{n=N_l}^{N_p}|\psi_n|^2}, \qquad (6)$$

where the summations at the denominator have been computed at the injection time t=0, and those at the numerator only after the time interval, $(N_p+N_l)/v_p$, the transmitted wave-packet takes to reach the midpoint of chain on the opposite side of a ring with ε=g=0. In our simulations the residual norm trapped inside the ring was negligible for any choice of $\varepsilon_\pm$ and $g_\pm$.

With regard to the denominators in equation (6), we notice that, while the normalization constant of the wave-packet at time t=0, equation (3), has been set $I_0$=0.6 throughout our simulation work, the results reported in Figures 1-4 hold also for different normalization constants, $I_0'=\kappa I_0$, provided the nonlinearity constant $g_0$ is rescaled as $g_0 \rightarrow g_0' = \kappa^2 g_0$. In particular, the wave-packet of equation (3) can be normalized to one incident particle by setting κ=0.15.

The probability spectral densities (p.s.d) at site n=0 were computed by means of the Fast Fourier Transform (FFT) utility of the software package OriginPro8. The time series of the wave function $\psi_0(t)$ used in our computations were recorded between the time when the incident wave packet hit the injection ring site and the time when the center of the transmitted wave packet reached the midpoint of the exit chain.

**Model symmetries.** The symmetry relations of equation (5) have been demonstrated numerically, but could well have been predicted on the basis of simple analytical arguments.

1. $R_f(-\Phi)=-R_f(\Phi)$ follows from the identity $T_{LR}(\Phi)=T_{RL}(-\Phi)$, see Figure 2a, which is due to the right-left mirror symmetry of the device with respect to its center. Moreover, replacing n by –n in equation (1), however, inverts the sign of Φ.

2. $R_f(-k_0)=-R_f(k_0)$ follows from the very definition of $R_f$, equation (4), and the observation that by inverting the sign of $k_0$ one also interchanges the diode injection and exit site.

3. $R_f(-\varepsilon,-g)=-R_f(\varepsilon,g)$ is a property of the cubic DNSE. Equation (1) describes the same transport dynamics even if one takes its complex conjugate after substituting $\psi_n$ by $\psi_n^*$. Such a transformation would correspond to inverting the signs of $\varepsilon_n$, $g_n$, and taking the complex conjugate of $t_{m+n,n}$. Taking the complex conjugate of the hopping constant can be restored by replacing $k_0$ with $-k_0$. Numerical evidence of this symmetry relation is not reported in this work.

**Acknowledgements**

We thank Prof. Nianbei Li for useful discussions. This work is supported by the NSF China, with grant No. 11334007. Y. L is also supported by Shanghai Rising-Star Program with grant No. 13QA1403600.

**Additional information**

Competing financial interests: The authors declare no competing financial interests.

**Author contributions**

Y. L. carried out the numerical simulations and data analysis. J. Z. contributed to elaborate the model, F. M. wrote the main manuscript text, B. L. attended the discussions and supervised the project. All the authors contributed to interpret the results and review the manuscript.

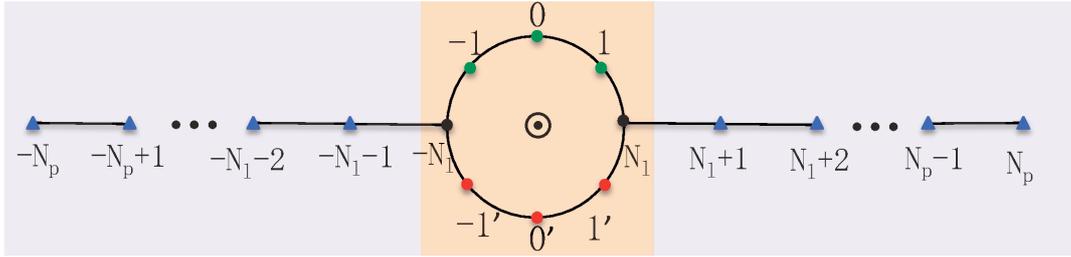

Figure 1| Sketch of an AB wave diode: an upside-down asymmetric, nonlinear ring-shaped lattice is connected to two elastic chains with $N_p$ = 500 and $N_l$ = 2. The magnetic field is applied perpendicularly to the ring. The ring site interactions have parameters $\varepsilon_\pm=\varepsilon_0(1\pm\delta_\varepsilon)$ and $g_\pm=g_0(1\pm\delta_g)$, see equation (1), where ± denote the upper (lower) half ring, $\delta_\varepsilon$ and $\delta_g$ control their asymmetry, and $g_0$ sets the degree of nonlinearity in the ring. At the junction sites, n=$N_l$, $\delta_\varepsilon=\delta_g=0$.

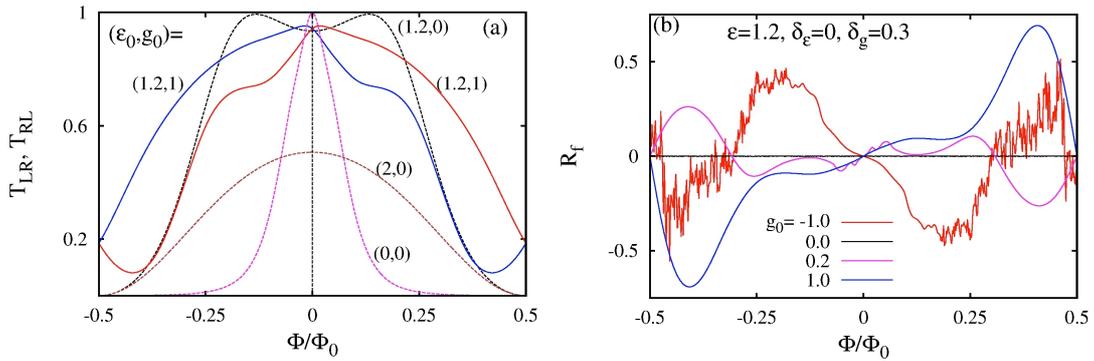

Figure 2| (a) Dependence of the transmission coefficients, $T_{LR}$ (solid blue) and $T_{RL}$ (solid red), on the magnetic flux Φ in units of $\Phi_0$. The parameters $\varepsilon_0$, and $g_0$ are reported next to each curve and $\delta_g$=0.3. The dashed curves represent $T_{LR}=T_{RL}$ as obtained for $g_0$=0, $\delta_\varepsilon$=0 and different $\varepsilon_0$. (b) Rectification factor, $R_f$ vs. $\Phi/\Phi_0$, for different $g_0$ (see legend). The simulation parameters are $g_0$=1, $\varepsilon_0$=1.2, $\delta_g$=0.3 and $\delta_\varepsilon$=0. In both panels the parameters of the wave-packet are $I_0$=0.6, $n_0$=250, $\sigma_0$=25 and $k_0=\pi/2$, see equation (3), and the ring is as in Figure 1.

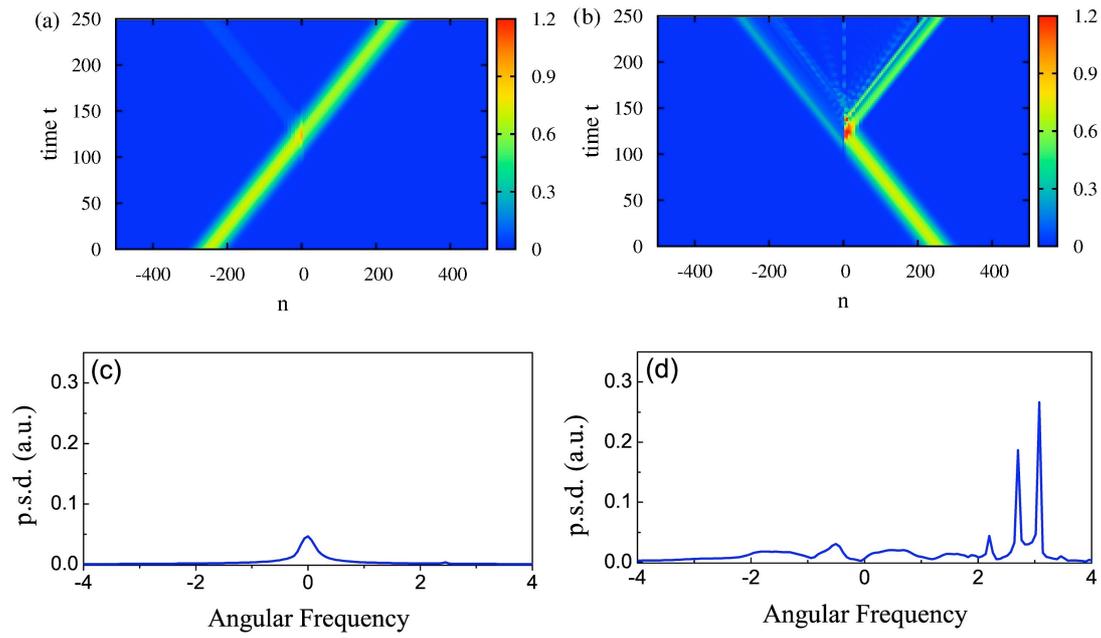

Figure 3| Upper panels: Time evolution of the Gaussian wave-packet of Figure 2 injected respectively from the left (a) and right side (b) of the ring. The norm of the wave at each site of the device is represented according to the color code bar on the r.h.s. of the panels. Here, $g_0=1$, $\Phi/\Phi_0=-0.2$ and all remaining simulation parameters are as in Figure 2b. Lower panels: Corresponding probability spectral densities (p.s.d.) at site $n=0$ in the same arbitrary units (see Methods).

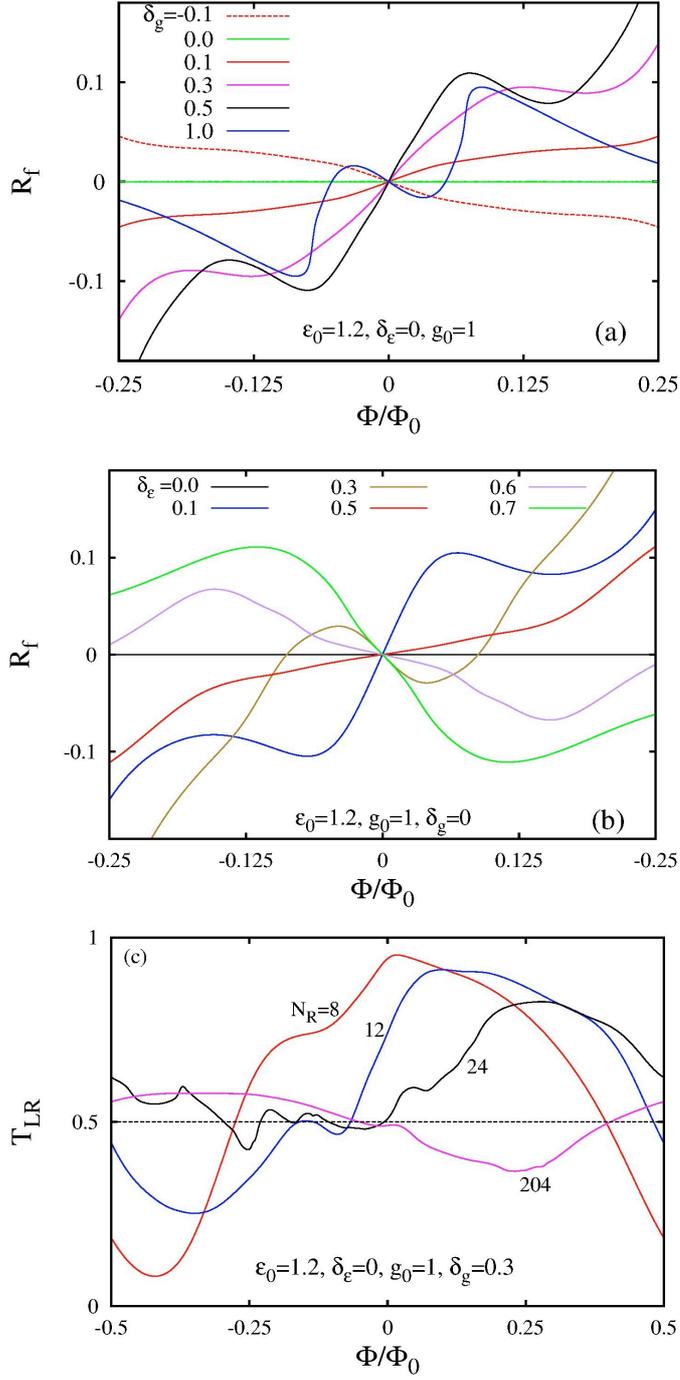

Figure 4| Diode performance on varying asymmetry and size of the AB ring: $R_f$ vs. $\Phi/\Phi_0$ for (a) different $\delta_g$ but $\delta_\varepsilon=0$ and (b) different $\delta_\varepsilon$ but $\delta_g=0$; (c) $T_{LR}$ vs. $\Phi/\Phi_0$ for different ring circumferences $N_R$. In all panels $\varepsilon_0=1.2$, $g_0=1$ and the injected wave-packet is as in Figure 2. Moreover, in (a) and (b) $N_R=8$, and in (c) $\delta_g=0.3$ and $\delta_\varepsilon=0$.